# Tracing the fate of carbon and the atmospheric evolution of Mars

Renyu Hu[1,2], David M. Kass[1], Bethany L. Ehlmann[1,2] & Yuk L. Yung[1,2]

The climate of Mars likely evolved from a warmer, wetter early state to the cold, arid current state. However, no solutions for this evolution have previously been found to satisfy the observed geological features and isotopic measurements of the atmosphere. Here we show that a family of solutions exist, invoking no missing reservoirs or loss processes. Escape of carbon via CO photodissociation and sputtering enriches heavy carbon ($^{13}$C) in the Martian atmosphere, partially compensated by moderate carbonate precipitation. The current atmospheric $^{13}$C/$^{12}$C and rock and soil carbonate measurements indicate an early atmosphere with a surface pressure <1 bar. Only scenarios with large amounts of carbonate formation in open lakes permit higher values up to 1.8 bar. The evolutionary scenarios are fully testable with data from the MAVEN mission and further studies of the isotopic composition of carbonate in the Martian rock record through time.

[1] Jet Propulsion Laboratory, California Institute of Technology, Pasadena, California 91109, USA. [2] Division of Geological and Planetary Sciences, California Institute of Technology, Pasadena, California 91125, USA. Correspondence and requests for materials should be addressed to R.H. (email: renyu.hu@jpl.nasa.gov).

The evolution of the atmosphere of Mars is one of the most intriguing problems in the exploration of the solar system. Presently Mars has a very thin 6-mbar atmosphere in equilibrium with polar caps and regolith. Yet, both morphological and mineralogical evidence suggests that the climate of Mars more than 3 billion years ago was warmer and wetter than the present[1]. The atmospheric conditions conducive to this environment are still uncertain. A denser $CO_2$ atmosphere may have facilitated early warm and wet surface conditions, at least locally and transiently at high orbital obliquities[2].

The pressure of the early Martian atmosphere has not yet been constrained by models of atmospheric evolution, due to uncertainties in the planet's early outgassing history, atmospheric escape and carbonate precipitation[3,4]. To transition from a thicker early atmosphere to the thin current atmosphere, carbon needs to be removed by either escape to space[5,6] or deposition near the surface as carbonates[7]. Recent models of the upper atmosphere of Mars suggest that <300 mbar of $CO_2$ has escaped to space since the late heavy bombardment (LHB)[8,9], and current Mars exploration has only found local evidence of carbonate deposits[10]. Neither mechanism, alone or coupled, fully accounts for the 'missing' $CO_2$, if a multi-bar early atmosphere is assumed.

Here we combine the recent Mars Science Laboratory (MSL) isotopic measurements of Mars' atmosphere, and orbital remote sensing and in situ measurements of Mars' surface composition to place hard constraints on Mars' atmospheric evolution. The isotopic signature of carbon offers a unique tracer for the atmospheric evolution of Mars because $CO_2$ is the major constituent of Mars' atmosphere, and because carbon is not incorporated into surface minerals except for carbonates. Our study is driven by the following three recent and important constraints from in situ and remote sensing observations.

**Carbon isotope signature of Mars' atmosphere.** Early data analyses from the Phoenix lander showed an isotopically light atmosphere but were influenced by a calibration artefact[11,12]. Telescopic measurements had a large uncertainty of 20‰ and were subject to telluric contamination[13]. The Sample Analysis at Mars (SAM) instrument suite on MSL has reported the most precise isotopic measurements of atmospheric $CO_2$ to date: $\delta^{13}C = 46 \pm 4$ (ref. 14), measured by both the tunable laser spectrometer and the quadrupole mass spectrometer, which shows that the current Martian atmosphere is enriched in $^{13}C$ than the Martian mantle (Fig. 1). $\delta^{13}C$ is defined as the relative enhancement of the ratio $^{13}C/^{12}C$ with respect to a reference standard (VPDB), reported in parts per thousand (‰):

$$\delta^{13}C \equiv \frac{(^{13}C/^{12}C)_{Sample} - (^{13}C/^{12}C)_{VPDB}}{(^{13}C/^{12}C)_{VPDB}} \times 1,000 \quad (1)$$

where $(^{13}C/^{12}C)_{VPDB} = 0.0112372$ (ref. 15).

**Carbonates formed in the Amazonian era.** Orbital remote sensing indicates the Martian dust contains 2–5 wt% of carbonate[16]. Phoenix-evolved gas experiments have measured up to 6 wt% carbonate in soil of the northern plains[17]. MSL's evolved gas experiments in Gale Crater found ~1 wt% carbonate at the Rocknest aeolian deposit[18]. Certain young, large geologic units on Mars, including the southern polar layered deposits and the Medusae Fossae formation may contain up to 10 m global equivalent of dust[19,20]. On the basis of these measurements, we estimate an upper limit of carbonate formation during the Amazonian Era to be 7 mbar of $CO_2$, corresponding to global presence in the upper 10 m of soil at an abundance of ~2 wt%.

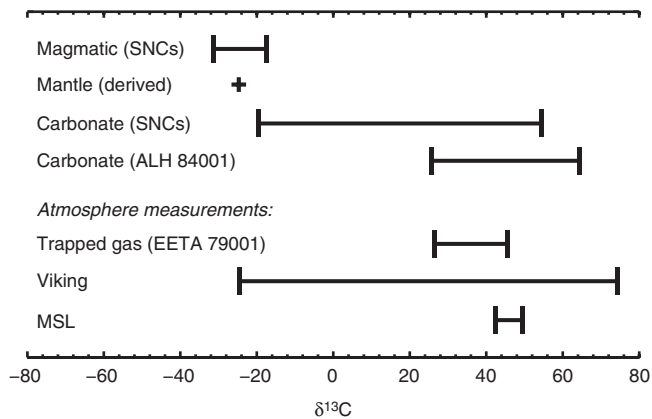

**Figure 1 | Summary of $\delta^{13}C$ measurements of Mars.** The $\delta^{13}C$ value of the magmatic component of SNC meteorites are used to derive the $\delta^{13}C$ value of the Martian mantle[31]. The carbonates in SNC meteorites have highly variable $\delta^{13}C$ values and the carbonates in ALH 84001 (formed ~3.9 Ga before the present) are generally enriched in $^{13}C$ than the Martian mantle[11]. The rest of the measurements are for modern Mars' atmosphere[14,51,52].

**Carbonates formed in the Noachian and Hesperian era.** Carbonate-bearing rocks have been discovered in various Noachian terrains by orbital remote sensing[10] and in one rock formation at Gusev Crater in situ[21]. The largest contiguous exposure of carbonate-bearing rocks in Nili Fossae covers 15,000 km$^2$, is a few tens of metres thick, and may host up to 12 mbar of $CO_2$ (refs 22,23). Deep crustal carbonate rocks may also exist, exposed in several impact craters[24]. However, carbonates are not widespread on Mars and are rare compared with other secondary minerals like hydrated silicates and sulfates[25]. There is no inherent difference in the detectability between phyllosilicate and carbonate from an infrared spectroscopy methodological perspective, and the planet Mars has been globally sampled. Thus, the difference is most simply explained as a real difference in abundance. On the basis of this fact, an upper bound of the amount of carbonates is 5 wt% in the volume of crust interrogated by remote sensing, constrained by an upper bound of carbonate non-detectability from infrared absorption features. Assuming 500 m depth, this upper bound corresponds to an equivalent atmospheric pressure of 1.4 bar. Similarly, 1 wt% of carbonates in the crust, more plausibly non-detectable in remote sensing and rover-based analyses, would correspond to 0.3 bar of $CO_2$.

Driven by these three constraints and together with a newly identified mechanism (photodissociation of CO) that efficiently enriches the heavy carbon isotope, we find a group of plausible atmospheric evolution solutions that can indeed satisfy the current atmospheric pressure and isotopic signatures, and the amount of carbonate deposition, invoking no missing reservoirs or loss processes. We therefore derive new quantitative constraints on the atmospheric pressure of Mars through time, extending into the Noachian, ~3.8 Gyr before the present. The atmospheric $\delta^{13}C$ data and the carbonate content in rock and soil indicate an early atmosphere with a surface pressure <1 bar. Only scenarios with large amounts of carbonate deposition in open-water systems permit higher values up to 1.8 bar.

## Results

**Isotope fractionation in CO photodissociation.** Photodissociation of CO is the most important photochemical source of escaping carbon atoms from Mars, responsible for ~90% photochemical loss[9]. Its fractionation factor, however, has never been evaluated. Here we show that CO photodissociation on Mars has a fractionation factor of 0.6 and is a highly efficient mechanism to enrich $^{13}C$ of the atmosphere.

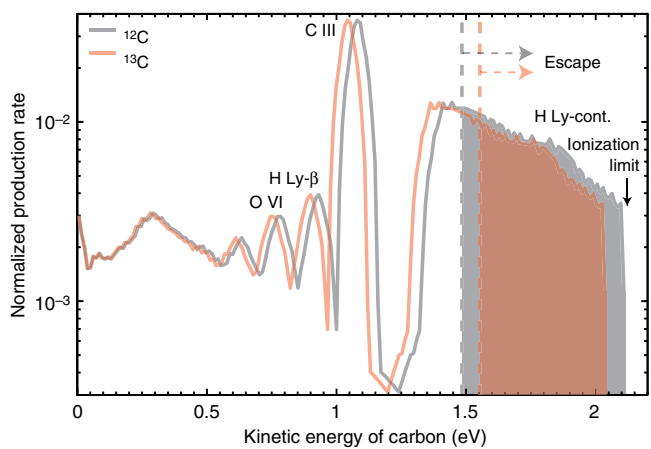

**Figure 2 | Energy distribution of carbon atoms produced by photodissociation of CO.** The critical energy for each isotope to escape is shown in dashed lines for comparison. The energy distribution is calculated using the current solar minimum spectrum for an exobase at 200 km and the measured CO photodissociation cross-section[26]. The grey and red areas indicate the fraction of $^{12}C$ and $^{13}C$ that escape, respectively, which is 0.40 for $^{12}C$ and 0.24 for $^{13}C$. The fractionation ratio $^{12}C/^{13}C$ via CO photodissociation is thus 0.6. Using early Sun proxies[47] or assuming higher exobases gives quantitatively similar results.

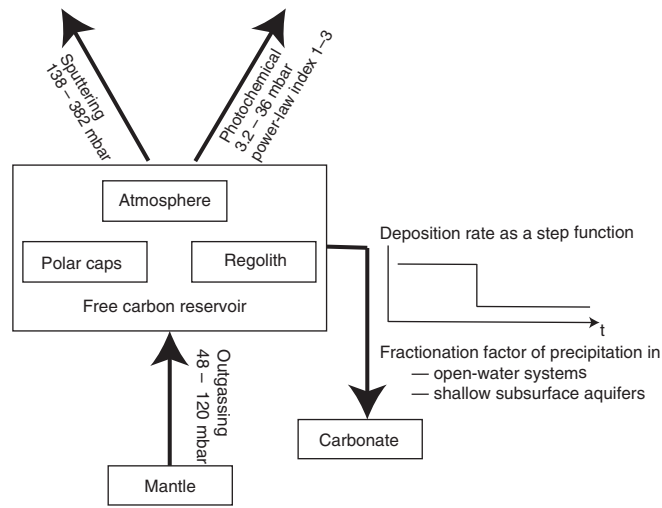

**Figure 3 | A box model for long-term exchanges between the carbon reservoirs on Mars.** The model traces a single free-carbon reservoir that includes the atmosphere, $CO_2$ in polar ice, and adsorbed $CO_2$ in the regolith, with magmatic outgassing as the source, and carbonate deposition and atmospheric escape as the sinks. The escape mechanisms include pick-up ion sputtering and photochemical escape.

In a CO photodissociation, energy from the incident photon, in excess of the bond dissociation energy, is imparted to carbon and oxygen atoms as kinetic energy. We use the solar spectrum and the cross-section of CO photodissociation as a function of wavelength to calculate the kinetic energy distribution of carbon isotopes (Fig. 2). The significant fractionation effect of CO photodissociation is mainly due to two effects: first, conservation of momentum determines that $^{13}C$ takes a lesser fraction of the excess energy than $^{12}C$ in each photodissociation event; and second, $^{13}C$ requires more energy to escape from the gravity of Mars. The excess energy needs to be >2.6–2.9 eV to produce escaping carbon atoms (that is, the escape threshold energy is 1.5 eV for $^{12}C$ and 1.6 eV for $^{13}C$, and the corresponding energy for non-escaping $^{16}O$ by conservation of momentum is 1.1–1.3 eV). Because the bond dissociation energy of CO is 11.2 eV, escaping carbon atoms can only be produced by photons more energetic than 13.8 eV, that is, the solar Lyman continuum. In this regime, the cross-section of CO photodissociation does not have strong lines[26]. Furthermore, the branching ratios of CO photodissociation do not affect the fractionation factor, because only the channel that produces ground-state atoms leads to escape of any carbon atoms. If one of the dissociation products is in its excited state (for example, $C(^1D)$ or $O(^1D)$), the produced carbon atom will not have enough kinetic energy to escape for any photon less energetic than the ionization threshold (83.5 nm). Thus, the fractionation factor of 0.6 is not sensitive to the evolution of the solar extreme ultraviolet (EUV) spectrum.

**Martian atmosphere evolution scenarios.** With the newly calculated fractionation factor for CO photodissociation, we construct a model to trace the history of $\delta^{13}C$ of a free-carbon reservoir with carbonate deposition and atmospheric escape as the two sinks and magmatic activity as the sole source (Fig. 3).

The free-carbon reservoir includes all reservoirs exchangeable on short timescales, that is, the atmosphere, $CO_2$ in polar ice and adsorbed $CO_2$ in the regolith. The carbon history of Mars has been extensively studied with reservoir models[3–7,27–30], but none of the previous models have included photochemical escape as a major mechanism that enriches atmospheric $^{13}C$.

We model the evolution of this reservoir starting from 3.8 Ga before the present, that is, the mid/late Noachian after any impact-enhanced loss during the LHB, beginning with an atmospheric $\delta^{13}C$ value equal to that of mantle-degassed $CO_2$ derived from the magmatic component of the SNC meteorites (shergottites, nakhlites, chassignites)[31] (other starting values produce similar results; see the next section). The modelled initial reservoir size is calculated from the sum of the current reservoir size, the total removal by atmospheric escape and the total amount of carbonate formation, minus the total outgassed. For outgassing, we adopt the volcanic emplacement rates from thermal evolution models and photogeological estimates, which are in agreement (48–120 mbar)[32,33]. For sputtering, we adopt the three-dimensional (3D) Monte Carlo simulations[8] as the standard escape rates, and the total sputtered is 138–382 mbar, depending on the age dependency of the solar flux. We adopt the photochemical escape rate calculated for the present-day solar flux[9], and scale up the rate for earlier, more intense solar Lyman continuum with a power law:

$$F_{\mathrm{pr}} = 7.9 \times 10^{23} F_{\mathrm{Lyc}}^{a} \qquad (2)$$

where $F_{\mathrm{pr}}$ is the photochemical escape flux in particles per s, $a$ is the power-law index and $F_{\mathrm{Lyc}}$ is the solar Lyman continuum flux in units of the current solar Lyman continuum flux. The power-law index is a free parameter in the model, because existing calculations of the photochemical escape rate only study current solar conditions[9,34]. It is however reasonable to explore a power-law index ranging between 1 and 3, because an increasing solar EUV flux would lead to increasing mixing ratios of CO, $CO^+$ and electrons in the thermosphere, and these multiple factors could contribute to increasing the photochemical escape rate of carbon. The total photochemical loss for this range is 3.2–36 mbar.

The rate of carbonate formation is simply assumed to be a step function, characterized by an early carbonate formation rate, a late carbonate formation rate and a time of transition. Reality would, of course, be a more gradual transition. For the effects on the carbon isotopic ratio of the atmosphere, two carbonate formation scenarios are considered. One scenario is precipitation in open-water systems (for example, lake and ponds) that have good isotopic communication with the atmosphere. The carbonate formed in this way is ~10‰ enriched than the atmosphere[15]. The other scenario is precipitation in shallow subsurface aquifers that are semi-isolated and have poor isotopic communication with the atmosphere, that is, there is no influx of gas to replace carbonate precipitated in rock pores. Carbonate is formed by evaporation of water originally derived from the surface, and the water can be enriched by up to ~50‰ after 99% of the original volume evaporated. The carbonate formed in this way is thus up to ~60‰ enriched relative to the atmosphere. The shallow subsurface aquifer scenario has been suggested to explain the high $\delta^{13}C$ values of the carbonates in the Martian meteorite ALH 84001 (ref. 35). Some carbonate formation may also proceed in subsurface, closed aquifers from carbon-bearing gases sourced from hydrothermal fluids, but these would not influence the atmospheric reservoir's evolution.

We undertake a million-model approach to quantify the relationship between the amount of carbonate formation and the escape rate, and derive constraints on the early surface pressure. We explore the power-law index of the photochemical escape rate from 1 to 3, the time of transition from high carbonate formation to low carbonate formation ranging from 3.0 to 3.5 Ga, the amount of early carbonate formation from 0.001 to 10 bar, the amount of late carbonate formation from 0.01 to 7 mbar and the amount of early volcanic outgassing from 48 to 120 mbar. We also consider the uncertainties in how the solar EUV flux varies with age, which affect the total removal of sputtering and photochemical processes (see Methods). We performed ~50 million simulations using combinations of parameters for both carbonate formation scenarios, and show the combinations of parameters that produce the $\delta^{13}C$ value in the 1–$\sigma$ range measured by MSL in Fig. 4. The range of early surface pressure—including the atmosphere, the absorbed carbon in the regolith and the polar caps—is also shown in Fig. 4.

Most scenarios permitted by the measurements of the current atmospheric $\delta^{13}C$ and the surface carbonate content have an early surface pressure <1 bar. Owing to the $^{13}C$ enrichment effect, even a small amount of atmospheric escape via CO photo-dissociation can drive the atmospheric $\delta^{13}C$ value to the present-day value measured by MSL. Therefore, the isotopic data themselves do not require massive atmospheric loss, and existing known escape mechanisms are fully consistent with all evidence from measured isotopic values and carbonate abundance. In fact, the enrichment from 3.8 billion years ago to present is so significant that it must be compensated by Noachian/Hesperian carbonate deposition, because carbonate formation and outgassing during the Amazonian are low (see the next section for a discussion on Amazonian volcanic outgassing). Figure 4 quantifies this compensation: a higher photochemical escape flux implies a greater amount of early carbonate deposition; but if more carbonates precipitated in shallow subsurface aquifers that were locally enriched, a lesser volume would be required. If all carbonate precipitated in highly enriched shallow subsurface aquifers as ALH 84001, the upper bound of carbonate formation is 0.5 bar, yielding an upper bound of the surface pressure of 0.9 bar. A surface pressure >1 bar is only permitted if the power-law index is >2 and most carbonates formed in open-water systems. That would also require more carbonates, not yet detected by rovers and orbiters. The upper bound on the amount of carbonates allowed by the geologic record (5 wt% everywhere globally in the top 500 m, or 1.4 bars of $CO_2$) thus determines an overall maximum early atmospheric pressure of 1.8 bar. The two carbonate formation scenarios examined are endmembers, and any solution between the two is viable and results in initial atmosphere values between the two cases.

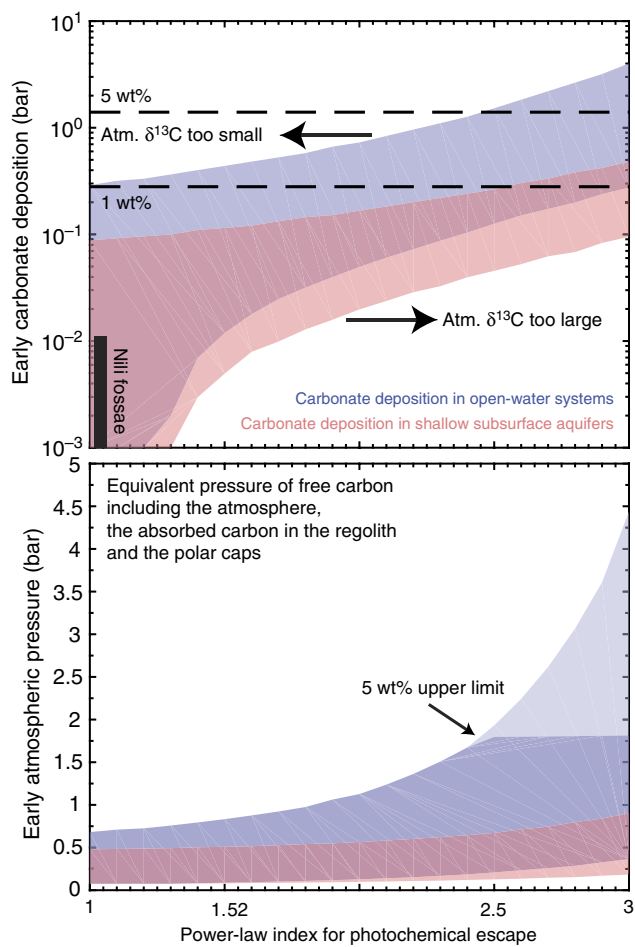

**Figure 4 | Carbonate formation and early surface pressure.** These are constrained by the current atmospheric $\delta^{13}C$ value measured by the MSL[14]. In both panels the red area shows the permitted range when carbonate deposition occurred in shallow subsurface aquifers and the blue area shows the permitted range when carbonate deposition occurred in open-lake systems. The amount of early carbonate formation must be commensurate with the amount of photochemical escape to produce the measured $\delta^{13}C$ of the current Mars' atmosphere. A lesser amount of carbonate formation is required if the fractionation factor is greater. The early surface pressure is constrained by both the current atmospheric (Atm.) $\delta^{13}C$ value and the upper limit of 1.4 bar (5 wt%) for the early carbonate formation, shown in dark blue versus light blue colour in the bottom panel.

Figure 5 shows four standard scenarios that lead to a present-day $\delta^{13}C$ value consistent with the MSL measurement. The scenarios are chosen for a power-law index of 2 for the photochemical escape rate. Calculations of the photochemical escape rate for present-day low solar activity and high solar activity indicate a power-law index of 2–2.4 (refs 9,34), although the range of the EUV flux in these calculation (a factor of 2) is smaller than the range from the present day to 3.8 Ga. Two observations can be made from the evolutionary tracks of these scenarios. First, photochemical processes are the main processes that enrich $^{13}C$, although sputtering is the main atmospheric escape process. The amount removed by sputtering is ~30 times

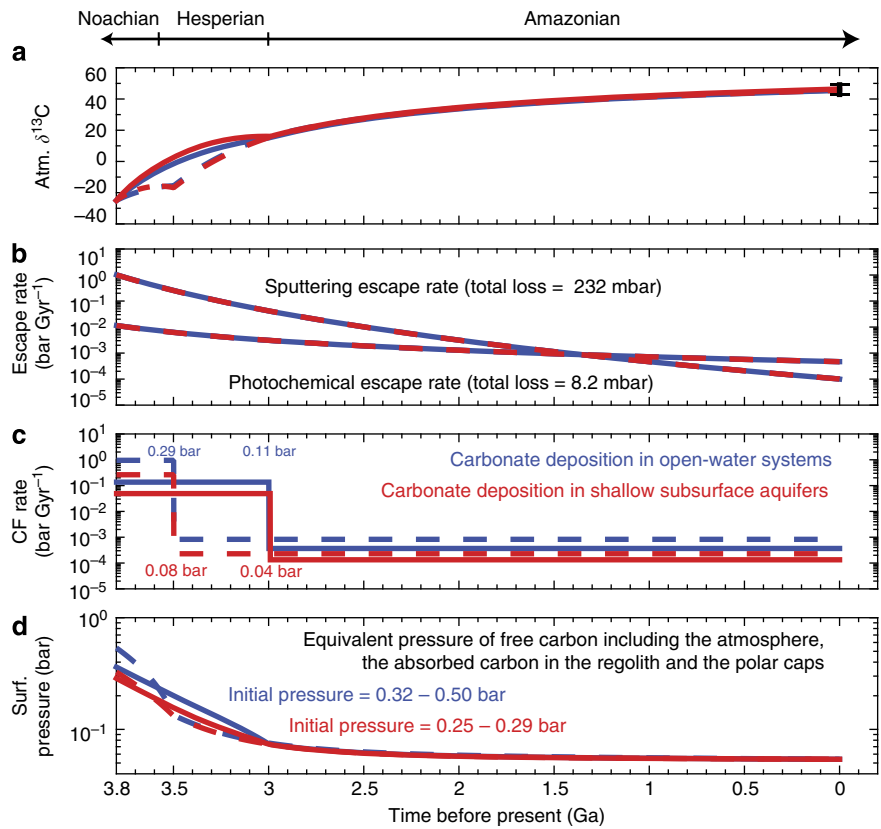

**Figure 5 | Standard scenarios of carbon evolution on Mars since the LHB at 3.8 Ga that arrive at present-day $\delta^{13}C$ values.** (**a**) The evolution of the atmospheric (Atm.) $\delta^{13}C$ value, in comparison with the MSL measurement shown by the error bar. (**b**,**c**) The evolution of the escape rate and the carbonate formation rate, respectively. All scenarios have the same escape rate, corresponding to a power-law index of 2 for the photochemical escape rate. The solid lines are the scenarios where carbonate deposition persisted through the Hesperian Era and the broken lines are the scenarios where carbonate deposition only occurred during the Noachian Era. The blue lines are the scenarios where carbonate deposition occurred in shallow subsurface aquifers and the red lines are the scenarios where carbonate deposition occurred in open-water systems. For each scenario, the amount of early carbonate formation (CF) is determined by the $\delta^{13}C$ value. (**d**) The evolution of the surface (Surf.) pressure. The definition of the age boundaries is taken from a recent crater-density study[53]. The escape rate in the second panel can be converted to atom per s by 1 bar Gyr$^{-1}$ = $1.7 \times 10^{27}$ atom per s.

**Table 1 | Jacobian values for how the amount of sputtering escape ($M_{SP}$), photochemical escape ($M_{PH}$), carbonate deposition ($M_{CD}$) and volcanic outgassing ($M_{VO}$) would affect the present $\delta^{13}C$.**

| Early carb. (bar) | Late carb. (bar) | Transition time (Ga) | Index | $\partial\delta^{13}C/\partial\ln(M_{SP})$ | $\partial\delta^{13}C/\partial\ln(M_{PH})$ | $\partial\delta^{13}C/\partial\ln(M_{CD})$ | $\partial\delta^{13}C/\partial\ln(M_{VO})$ |
|---|---|---|---|---|---|---|---|
| *Carbonate deposition in open-water systems* | | | | | | | |
| 0.29 | <0.007 | 3.5 | 2.0 | 74.0 | 78.9 | −46.3 | −0.34 |
| 0.11 | <0.007 | 3.0 | 2.0 | 66.3 | 65.1 | −54.2 | −0.46 |
| *Carbonate deposition in shallow subsurface aquifers* | | | | | | | |
| 0.08 | <0.007 | 3.5 | 2.0 | 89.9 | 82.4 | −62.6 | −0.57 |
| 0.04 | <0.007 | 3.0 | 2.0 | 75.3 | 77.5 | −63.3 | −0.92 |

We determine the Jacobian values of each parameter by calculating the $\delta^{13}C$ for a variation of 10% of each parameter from the standard scenarios shown in Fig. 5. The columns 'Early carb' and 'Late carb' indicate the amount of carbonate deposition in the Noachian and Hesperian, and that in the Amazonian, respectively.

greater than the amount removed by photochemical processes, but their effects on the carbon isotopic ratio are comparable (Table 1). Second, if carbonate formation persisted through the Hesperian period, the required total amount of carbonates would be less than if carbonate formation only occurred during the Noachian. This is because a unit mass of carbonate formation has a greater impact on the final $\delta^{13}C$ value if it occurs later in the history. With the uncertainties in the time of transition and in the history of the solar EUV flux, the amount of $CO_2$ deposited as carbonate would be 20 mbar–0.7 bar, corresponding to 2–100 deposits of the size of Nili Fossae, or up to 3 wt%, if distributed globally. The early surface pressure is constrained to be 0.1–0.5 bar for carbonates formed in shallow subsurface aquifers, and the upper limit can be extended up to 1 bar for carbonates formed in open-water systems (Fig. 4).

Finally our results show that carbonate formation from the late Noachian to the Hesperian period is not required when the power-law index is <1.5 and the amount of sputtering is at the lower end of the reasonable range. These are the most 'carbonate conservative' scenarios fully consistent with the isotopic data, which do not require carbonate deposits beyond Nili Fossae and imply an early surface pressure of <0.3 bar.

**Sensitivity to the starting $\delta^{13}C$ and the outgassing history.** To understand the effects of the starting atmospheric $\delta^{13}C$ value and the volcanic outgassing rates to our results, we performed additional sets of simulations that assume the $\delta^{13}C$ value at 3.8 Ga before present to be −35 and −15‰, and simulations that assume higher volcanic outgassing rates or longer volcanic outgassing period than the standard scenarios. The results of these simulations are shown in Fig. 6.

If the starting atmosphere has a lower value for $\delta^{13}C$, fewer carbonate rocks would be required for a fixed photochemical escape rate. For example, for a power-law index of 2, the minimum amount of carbonate formation is 0.05 bar if the starting $\delta^{13}C$ is −25‰, 0.02 bar if the starting $\delta^{13}C$ is −35‰ and 0.09 bar if the starting $\delta^{13}C$ is −15‰ (the left panel of Fig. 6). The fractional variation of the required amount of carbonate deposition is significant when the carbonate deposition amount is small. When the amount of carbonate formation is >0.1 bar, we find the sensitivity to the initial $\delta^{13}C$ becomes much less significant. In general, the uncertainty in the amount of carbonate formation introduced by the starting $\delta^{13}C$ value is <0.1 bar (Fig. 6) and so is the uncertainty in the estimate of the early surface pressure.

We adopt the model by Grott et al.[32] for the outgassing rate history. Two endmember scenarios for outgassing are suggested, one is termed 'global melt' scenario and the other is termed 'mantle plume' scenario. The difference between the 'global melt' scenario and the 'mantle plume' scenario is that the latter has the volcanic outgassing flux more evenly distributed throughout the Hesperian period and extended into the Amazonian period (Fig. 7). For the standard models, we assume the 'global melt' scenario for the oxygen fugacity one order of magnitude higher than the iron-wustite buffer (IW+1) and an efficiency of $\eta = 0.4$ (for a total outgassing amount of 48 mbar). Increasing the efficiency to $\eta = 1$ (for a total outgassing amount of 120 mbar) results in quite minimal changes (Fig. 6).

Without changing the total outgassing rate, but more evenly distributing it over the Hesperian (that is, the 'mantle plume' scenario) would cause a decrease in the required amount of carbonate formation for a power-law index of ~1.5 (the right panel of Fig. 6). Prolonging the volcanic outgassing period decreases the minimum amount of required carbonate formation, because both processes lower the $\delta^{13}C$ value, and because outgassing during the late Hesperian has a greater impact on the final $\delta^{13}C$ value than that during the early Hesperian. To summarize, either increasing the total outgassing rate or prolonging the outgassing period lead to minor changes to our standard models, and our results appear to be relatively insensitive to the volcanic outgassing rates varied in a wide range.

For completeness, we test the evolution scenarios by both increasing the total outgassing rate and prolonging the outgassing activity. Specifically, we assume the mantle plume scenario of Grott for outgassing, at an oxygen fugacity of IW+1 and a degassing efficiency of 0.4. The total outgassing amount since 3.8 Ga would then be 420 mbar, in which 350 mbar would be outgassed between 3.8 and 3.0 Ga and 70 mbar would be outgassed between 3.0 Ga and present. This is compared with the standard models in which 48 mbar would be outgassed between 3.8 and 3.0 Ga, and essentially 0 would be outgassed after 3.0 Ga.

For this kind of volcanic outgassing history, the planet must have started to build-up the atmosphere from close to zero pressure at ~3 Ga before the present, to arrive at the appropriate present-day size of free carbon (that is, 54 mbar). This is a result of simple mass budget balance owing to the fairly significant volcanic outgassing source after 3.0 Ga and insignificant mass loss due to non-thermal escape. If the planet had an atmosphere at ~3 Ga, its current atmosphere would be more massive than 7 mbar.

To fit the current $\delta^{13}C$ value, the solution would be substantially different from our standard scenarios, in that the final $\delta^{13}C$ value is no longer sensitive to any evolutionary events before 3.0 Ga, including the early carbonate formation rate, and that the photochemical escape rate becomes the sole factor that controls atmospheric evolution since 3.0 Ga. In particular, we find that the power-law index would have to be >3.7 to provide enough fractionation during the Amazonian and lead to the present-day $\delta^{13}C$ value consistent with the MSL measurement. This solution cannot represent the evolution of planet Mars, because it requires Mars to have no or minimal atmosphere 3.0 Ga before the present, and an extremely large power-law index for the photochemical escape rate, both of which are unlikely. We therefore suggest that the solutions with substantial outgassing during the Amazonian period is unlikely for planet Mars.

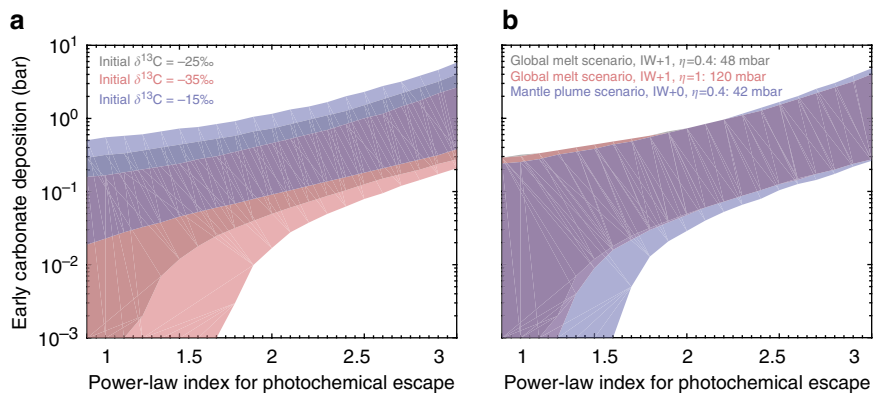

**Figure 6 | Sensitivity of the results to the initial $\delta^{13}C$ value and the volcanic outgassing rates.** The same as the upper panel of Fig. 4, but assuming $\delta^{13}C$ to be −35 and −15‰ at the beginning of the modelled period (**a**), or assuming different volcanic outgassing models (**b**). The coloured areas show the permitted range of the amount of carbonate deposition, assuming deposition in open-lake systems. Different colours correspond to different sensitivity studies, as labelled on the figure. The sensitivity results for the scenarios assuming deposition in shallow subsurface aquifers are similar. The constraints are relatively insensitive to the initial $\delta^{13}C$ value or the volcanic outgassing rates in reasonable ranges.

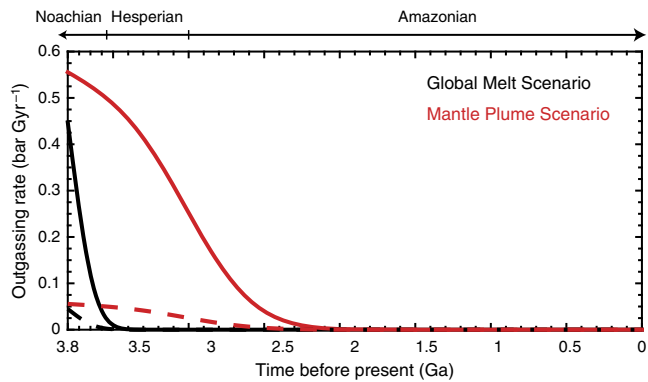

**Figure 7 | Volcanic outgassing rates.** We adopt the model by Grott et al.[32]. Cases of oxygen fugacity of IW + 1 and IW for an efficiency of 0.4 are shown in solid and dashed lines, respectively.

## Discussion

In this work we show that a large 'missing' carbon reservoir is unnecessary unless the volcanic outgassing rate was very high. Rather, starting from a few hundred mbar to about 1 bar of $CO_2$ after the LHB, sputtering and moderate carbonate formation is able to reduce the atmospheric pressure, and subsequent photochemical escape is able to appropriately enrich $^{13}C$ with known processes. The surface pressure constrained by isotopic modelling is also consistent with an upper bound provided by analyses of impact crater size[36]. The uncertainty in the surface pressure is dominated by the uncertainties in the photochemical and sputtering escape rates, as well as the geological settings of early carbonate formation. Our results thus highlight the crucial importance of a reliable understanding of non-thermal escape processes on Mars. How the photochemical escape rates scale with the solar EUV flux plays a key role, and is yet to be studied. The Mars Atmosphere and Volatile EvolutioN (MAVEN) mission will provide data to calibrate current non-thermal escape models and improve the extrapolation to early Mars.

The evolutionary tracks shown in this paper can be compared with carbonates in SNC meteorites that formed in the past 1 billion years and have varied $\delta^{13}C$ (ref. 10). Their formation from atmosphere-sourced carbon, mixed with magma-sourced carbon, is fully consistent with our evolution scenarios. Importantly, our model provides a methodology for determining even more precisely the past atmospheric escape flux, testable by examination of isotopic ratios in Martian carbonates. Coupled measurements of isotopic ratios and isotopologues of Noachian to Early Amazonian deposits, measured *in situ* or in meteorites and returned samples, would uniquely distinguish the timing of major carbonate formation, its geologic setting, as well as the amount of carbonate deposition that balanced photochemical escape.

## Methods

**A box model for the atmospheric evolution.** We adopt the box model of Kass[37] to trace the evolution of the free-carbon reservoir and its $\delta^{13}C$ value. We make major changes to the model, including: using the latest calculations of sputtering rates; adding photochemical escape as a mechanism of non-thermal escape and calculating the fractionation factor of CO photodissociation; using the latest estimate of the outgassing history of Mars that has taken into account the newly measured solubility of carbon in a more reduced mantle melt; considering the atmosphere, polar caps and the regolith to be one exchangeable reservoir for free carbon, and using the latest measurements of their masses for the current size; and modifying the implementation of the atmospheric collapse according to recent 3D early Mars climate simulations and allowing the model to be directly constrained by the geologic record.

**Free-carbon reservoir.** The current Martian atmosphere is at or near equilibrium with polar caps and regolith. We define 'free carbon' as the atmosphere, $CO_2$ ice in polar caps and $CO_2$ absorbed by the regolith. All free-carbon species are treated as a single reservoir, assuming they exchange with each other over geological times and they do not significantly fractionate relative to each other in this exchange. Laboratory measurements indicate that the carbon in $CO_2$ does not fractionate during condensation, an unusual phenomenon probably due to the effect of isotopic substitution on the inter-molecular binding energy of the condensed phase[38]. Recent measurements also show that regolith absorption results in a ~1‰ enrichment of $^{13}C$ in the gaseous phase[39]. Therefore, condensation or absorption of even large amounts of $CO_2$ does not change $\delta^{13}C$ in the atmosphere.

Detailed characterization of Mars and relevant laboratory studies in recent years have provided good constraints on the size of this free-carbon reservoir. We adopt a current free-carbon reservoir of 54 mbar (7 mbar atmosphere, 2 mbar polar caps[40], 5 mbar subsurface polar deposits[41] and 40 mbar regolith absorption[42]) as a fixed parameter in our baseline models.

The advantage of combining these reservoirs into a single reservoir is that we do not have to explicitly trace the evolutionary history of the surface temperature, which is primarily a function of the surface pressure, the solar luminosity and orbital obliquity[2]. The surface temperature is the dominant factor that controls the partitioning of $CO_2$ among the atmosphere, the regolith and the polar caps[3]. This way, we can focus on evaluating the relationship between non-thermal escape that enriches $^{13}C$ and carbon deposition that depletes $^{13}C$.

**Outgassing.** Outgassing is the primary source of new carbon into the free-carbon reservoir. The carbon in the mantle is released into the atmosphere through volcanic emplacement of mantle material and outgassing from the magma. The outgassing flux can be estimated from the history of volcanic activity, the estimated intrusive emplacements and the carbon content of the magmas. The history of volcanic activity of Mars has been estimated photogeologically by determining the ages of volcanic units of the planet's surface[33] and theoretically by modelling the thermal history of the planet[32]. Photogeological estimates based on Viking data suggest extrusive magma of $67 \times 10^6$ km$^3$ from the late Noachian to present[33], which corresponds to a total outgassing of 0.5–50 mbar for complete degassing of 10–1,000 p.p.m. dissolved $CO_2$. In addition, for a ratio of intrusive to extrusive magma of 8.5:1, similar to Earth, the total outgassing can range from 2.4 to 240 mbar assuming 40% outgassing efficiency for the intrusive magma. No quantitative photogeological estimates have been published after, but recent observations using high-resolution imagery have suggested Martian volcanism started earlier and the decay in intensity was more rapid than previously thought[43].

The latest models of Mars' thermal evolution history appear to fully cover the uncertainties in the flux and timing of volcanic outgassing[32]. In the 'global melt' scenario of Grott, the planet cools fast, and most volcanic outgassing concentrates in the pre-Noachian and Noachian periods. For this scenario, the total outgassing rate during our modelled period (from 3.8 Ga to present) would be 48 mbar at an oxygen fugacity of IW + 1 and an outgassing efficiency of 40%. At IW + 1, ~1,000 p.p.m. $CO_2$ can be dissolved in the magma, and this amount scales with the oxygen fugacity exponentially[44]. In the 'mantle plume' scenario of Grott, the planet cools more slowly, and volcanic outgassing persists throughout the Hesperian period and extends into the Amazonian period. For this scenario, the total outgassing rate during our modelled period would be 470 mbar. The photogeological estimates are within the range defined by these two endmember scenarios.

For this study, we adopt the volcanic outgassing flux modelled by Grott, as

$$P_{\text{outgassing}} = 10^{\text{IW}} \eta A [\tanh(t/a)^\alpha]^{1/\beta} \quad (3)$$

where $P_{\text{outgassing}}$ is the cumulative outgassed partial pressure at $t$ Myr after formation and $\eta$ is the outgassing efficiency. For a 'global melt' scenario the parameters are $A = 252.45$ mbar, $a = 719.89$ Myr, $\alpha = 3.6206$ and $\beta = 6.7809$, and for a 'mantle plume' scenario the parameters are $A = 224.39$ mbar, $a = 1505.1$ Myr, $\alpha = 2.7606$ and $\beta = 3.3600$.

We assume volcanic outgassing has been mainly in the form of $CO_2$, rather than more reduced forms of carbon, during the modelled period. The speciation of volcanic outgassing on Mars is mainly controlled by the oxygen fugacity of the source magma. It has been recently shown, experimentally, that outgassing would be in $CO_2$ when the oxygen fugacity is $>$ IW $-0.55$ (ref. 45). Petrologically primitive SNC meteorites show that the Martian mantle has an oxygen fugacity between IW and IW + 1 (ref. 44). For this range, volcanic outgassing would be in $CO_2$. However, the early Martian mantle may have been more reduced than the sources of SNC meteorites, since the meteorite ALH 84001 formed 3.9 Ga before the present shows an oxygen fugacity as low as IW $-1$ (ref. 44). But, if this has been the case, the total outgassing rate would be insignificant (that is, $<12$ mbar for an oxygen fugacity of IW $-1$ for the most optimistic estimate of volcanic emplacement). We therefore suggest that the possibility of an early reduced mantle and consequently CO or even $CH_4$ outgassing during the modelled period has little impact on our model or results.

**Pick-up ion sputtering and photochemical escape.** During the model period, hydrodynamic escape has ceased, impact delivery and removal of volatiles are limited and the dominant atmospheric escape processes are pick-up ion sputtering and photochemical processes.

Pick-up ion sputtering is a process by which oxygen ions in Mars' upper atmosphere are picked up by the solar wind magnetic field and then collide with the molecules and atoms in the upper atmosphere sputtering them away[6,8]. This process may have been quite efficient in removing carbon from early Mars when the solar wind was much stronger than present. We expect sputtering occurred during the entire modelled period. Detection and mapping of crustal magnetic anomalies over the Mars surface implies the Martian magnetic field should have ceased before the formation of Hellas or the rise of Tharsis, because the interiors of these basins or most volcanic edifices lack magnetic remanence[46]. Therefore, the magnetic field does not affect our study. We adopt the 3D Monte Carlo simulations[8] as the standard values of the sputtering rate, fitted to this functional form

$$F_{sp} = \exp\left(-0.462\ln(F_{EUV})^2 + 5.086\ln(F_{EUV}) + 53.49\right) \quad (4)$$

where $F_{sp}$ is the sputtering escape flux in particles per s and $F_{EUV}$ is the solar EUV flux in units of the current solar EUV flux.

The evolution of the solar EUV flux has been derived by observing young solar-like stars[47]. We adopt $F_{EUV} \propto t^{-1.23 \pm 0.1}$, where $t$ is the age. The uncertainty of the index corresponds to 20% uncertainty in the flux at 3.8 Ga, determined from the measurement errors of the observations that are used to derive this flux index[47,48]. The total amount of atmosphere sputtered for the range of $F_{EUV} \propto t^{-1.23 \pm 0.1}$ is $232^{+150}_{-94}$ mbar, or 138–382 mbar, calculated using equation (4).

The carbon sputtered will be between 15 and 40‰ (depending on the epoch) lighter than the source atmosphere. The sputtering process itself does not fractionate the atmosphere, because the carbon atoms that actually escape all have sufficient energy to escape regardless of the isotope. But the sputtering occurs at altitudes well above the homopause[6,8], where each species takes on its mass-dependent scale height. Thus, the atmosphere being sputtered is lighter than the total atmosphere and the net effect of the sputtering is to enrich the atmosphere.

The fractionation factor α of the sputtering process can be calculated by

$$\alpha = \exp\left(\frac{-g\,\Delta m\,\Delta z}{kT}\right), \quad (5)$$

where $g$ is Mars' gravitational acceleration, $\Delta z$ is the distance from the homopause to the source altitude of escaping carbon atoms, $\Delta m$ is the mass difference between the isotopes, $k$ is the Boltzmann constant and $T$ is the mean temperature of the upper atmosphere. We adopt the source altitude from Kass[6] and calculate the fractionation factors to be 0.96–0.98. A later calculation appears to suggest a higher source altitude, by tens of kilometres[8]. The impact of this difference to the sputtering fractionation factor is minimal, as the fractionation factor of sputtering is close to unity in all cases, that is, sputtering is inefficient in enriching $^{13}C$ in the atmosphere compared with photochemical processes. However, it is efficient in removing atmospheric mass.

The main photochemical processes that generate escaping carbon atoms are photodissociation of CO and dissociative recombination of $CO^+$ and $CO_2^+$, and the escape rates are calculated for the present-day solar EUV conditions[9]. The escaping carbons are mainly produced by Lyman continuum photons (Fig. 2), and the appropriate way to scale up the rates with the solar Lyman continuum flux ($F_{Lyc}$) is unclear, so this is a free parameter in the model. We adopt $F_{Lyc} \propto t^{-0.86 \pm 0.1}$ based on observations of young solar-like stars[47,48]. Again, the uncertainty of the index corresponds to 20% uncertainty in the flux at 3.8 Ga, determined from the measurement errors of the observations used to derive this index[47,48]. The total amount of photochemical escape for the range of $F_{EUV} \propto t^{-0.86 \pm 0.1}$ is $3.4^{+0.4}_{-0.2}$ mbar for a power-law index of 1, $8.2^{+2.2}_{-1.6}$ mbar for a power-law index of 2 and $24^{+12}_{-8}$ mbar for a power-law index of 3, calculated using equation (2).

Photochemical processes fractionate the atmosphere in a much more efficient way than sputtering. This is because sputtering, when it occurs, is usually too energetic to have any isotopic effects, whereas the fate of photochemical products is sensitive to their masses. The fractionation factor of dissociative recombination for $^{13}C$ versus $^{12}C$ is ~0.8 (ref. 49), and that of CO photodissociation is ~0.6 (main text). Both depend weakly on the exobase location and fractionate more than sputtering (>0.95). Since about 90% of the photochemical escape flux is via CO photodissociation[9], the composite fractionation ratio is $0.8 \times 10\% + 0.6 \times 90\% = 0.62$.

**Carbonate deposition.** Since we do not explicitly trace the evolution of the atmospheric temperature and pressure, carbonate deposition has to be approximated. We assume the carbon deposition rate to be a step function, characterized by a constant relatively high early carbonate formation rate, a constant relatively low late carbonate formation rate and a transition time. We allow the transition time to vary between 3.5 to 3.0 Ga before the present, covering the scenarios where carbonate deposition persisted through the Hesperian Era, and the scenarios where carbonate deposition only occurred during the Noachian Era. Assuming the step function minimizes the number of free parameters and allows straightforward comparison with geologic evidence. In reality the carbonate formation rate would have been variable on small timescales due to transient existence of liquid water on the surface or in the subsurface. The cumulative amount of carbonate deposition during the Noachian and Hesperian, that during the Amazonian, and the transition time are three tuning parameters, and the first two are independently constrained by geologic record.

The standard temperature-dependent formulation of the carbon fractionation in carbonate formation indicates that carbonate precipitates in open-water systems at 0 °C is ~13‰ heavier than the source atmosphere. A temperature of 0 °C was selected because most of the carbonate rocks are expected to have formed in very cold water (since it is hard to warm the atmosphere even that much). However, the carbonates in the Martian meteorite ALH 84001 formed 3.9 Ga before the present and have high $^{13}C$ values[10]. One of the interpretations of the stable isotopic features of the carbonates of ALH 84001 is that they formed in an evaporative environment localized enriched in heavy carbon and oxygen, that is, shallow subsurface aquifers[35]. Therefore, we assume $\alpha_{carbonate} = \alpha_{precipitation} + \alpha_{evaporation}$, where $\alpha_{precipitation} = 13‰$ is the fractionation at precipitation and $\alpha_{evaporation}$ varies from 0 to 50‰, which corresponds evaporation of 0–99.2% of water before precipitation as calculated by the Rayleigh distillation formula. The upper bound of $\alpha_{evaporation}$ corresponds to ALH 84001. This treatment is coarse because the acidity of the residual solution will change as evaporation progresses[35]. Another possible way to increase $\alpha_{carbonate}$ above $\alpha_{precipitation}$ is that the source $CO_2$ was produced photochemically from a methane-rich background atmosphere[50]. For our purpose, it is not necessary to distinguish these possible causes for the large fractionation factor of carbonate formation suggested by ALH 84001.

### Acknowledgements

Support for this work was provided by NASA through Hubble Fellowship grant #51332 awarded by the Space Telescope Science Institute, which is operated by the Association of Universities for Research in Astronomy, Inc., for NASA, under contract NAS 5-26555. The research was carried out at the Jet Propulsion Laboratory, California Institute of Technology, under a contract with the National Aeronautics and Space Administration.

### Author contributions

R.H. modelled the fractionation factor of photochemical escape, developed the million-model approach, simulated the evolution scenarios and wrote the manuscript; D.M.K. provided the framework of the evolution model and modelled the fractionation factor of sputtering; B.L.E. assembled the measurements of carbonate content in rock and soil, and provided geological constraints on scenarios; Y.L.Y. provided the insight into the evolution of stellar radiation and escape rates; all authors interpreted the results and commented on the manuscript.

### Additional information

**Competing financial interests:** The authors declare no competing financial interests.

**Reprints and permission** information is available online at http://npg.nature.com/reprintsandpermissions/

**How to cite this article:** Hu, R. *et al.* Tracing the fate of carbon and the atmospheric evolution of Mars. *Nat. Commun.* 6:10003 doi: 10.1038/ncomms10003 (2015).